\documentstyle[aps,twocolumn,epsfig]{revtex}

\begin{document}

\title{Conducting phase in the \\ 
two-dimensional disordered Hubbard model}

\author{P.~J.~H. Denteneer} 
\address{Lorentz Institute, University of Leiden, P. O. Box 9506, 
2300 RA  Leiden, The Netherlands} 
\author{R.~T. Scalettar} 
\address{Physics Department, University of California, 1 Shields Avenue, 
Davis, CA 95616, USA}
\author{N. Trivedi} 
\address{Department of Theoretical Physics, Tata Institute of 
Fundamental Research, Homi Bhabha Road, Mumbai 400-005, India}

\maketitle

\begin{abstract}
We study the temperature-dependent conductivity $\sigma(T)$
and spin susceptibility $\chi(T)$
of the two-dimensional disordered Hubbard model.
Calculations of the current-current correlation function
using the Determinant Quantum Monte Carlo method show that 
repulsion between electrons can
significantly enhance the conductivity, and at low temperatures
change the sign of $d\sigma/dT$ from positive (insulating
behavior) to negative (conducting behavior). This result
suggests the possibility
of a metallic phase, and consequently a metal--insulator transition, 
in a two-dimensional microscopic model containing
both repulsive interactions and disorder.
The metallic phase is a non-Fermi liquid 
with local moments 
as deduced from a Curie-like temperature dependence of $\chi(T)$.

\end{abstract}



When electrons are confined to two spatial dimensions in a disordered
environment, common understanding until recently was that the electronic 
states would always be localized and the system would therefore be an 
insulator.
This idea is based on the scaling theory of localization \cite{gangof4}
for non-interacting electrons and corroborated by subsequent studies using
renormalization group (RG) methods\cite{Wegner}.
The scaling theory highlights the importance of the number of spatial
dimensions and demonstrates that while in three dimensions for 
non-interacting electrons there exists a transition from a metal to an 
Anderson insulator upon increasing the amount of disorder, a similar 
metal--insulator transition (MIT) is not possible in two dimensions. 

The inclusion of interactions into the theory has been problematic, 
certainly when both disorder and interactions 
are strong and perturbative approaches break down. 
Following the scaling theory the effect of weak interactions in the 
presence of weak disorder was studied by diagrammatic techniques
and found to increase the tendency to localize\cite{LR}.
Subsequent perturbative RG calculations, including both 
electron-electron interactions and disorder, found indications of
metallic behavior, but also, for the case without a magnetic 
field or magnetic impurities, found runaway flows to 
strong coupling outside the controlled perturbative regime
and therefore were not conclusive \cite{fink,BelKirk}.
The results of such approaches therefore have not changed the widely held 
opinion that in two dimensions (2D) the MIT does not occur.

The situation changed dramatically with the recent transport experiments 
on effectively 2D electron systems in silicon metal-oxide-semiconductor 
field-effect transistors (MOSFETs) which have provided surprising evidence 
that a MIT can indeed occur in 2D\cite{kravch}. 
In these experiments the temperature dependence of 
the conductivity $\sigma_{\rm dc}$ changes from that typical of an insulator 
(decrease of $\sigma_{\rm dc}$ upon lowering $T$) at lower density
to that typical of a conductor (increase of $\sigma_{\rm dc}$ upon 
lowering $T$) as the density is increased above a {\em critical} density. 
The fact that the data can be 
scaled onto two curves (one for the metal, one for the insulator)
is seen as evidence for the occurrence of a {\em quantum phase transition}
with carrier density $n$ as the tuning parameter.
The possibility of such a transition has stimulated a large number of  
further experimental \cite{popo,othexp} and also theoretical 
investigations \cite{GoF2,Chakra}, including proposals that
a superconducting state is involved \cite{superc}.
Explanations in terms of trapping of electrons at impurities, i.e. not
requiring a quantum phase transition have also been put forward\cite{AltKl}.
While there is no definitive explanation of the phenomena yet,
it is likely that electron-electron interactions play an important role.

The central question motivated by the experiments is:
Can electron-electron interactions enhance the conductivity of a 2D 
disordered electron system, and possibly lead to a conducting phase and 
a metal--insulator transition?
It is this question that we address by studying the 
{\em disordered Hubbard model}  
which contains both relevant ingredients: interactions and disorder. 
While the Hubbard model does not include the long range nature
of the Coulomb repulsion, studying the simpler model of screened
interactions is an important first step in answering the central
qualitative question posed above.
We use Quantum Monte Carlo simulation techniques which enable
us to avoid the limitations of perturbative approaches
(while of course being confronted with others).
We mention that recent studies using very different techniques from ours 
have indicated that interactions may enhance conductivity:
two interacting particles instead of one in a random potential has
a delocalizing effect \cite{TIP}, and weak Coulomb interactions were 
found to increase the conductance of spinless electrons in (small) 
strongly disordered systems \cite{Vojta}.


The disordered Hubbard model that we study is defined by: 
\begin{equation}
{\hat H} = - \sum_{i,j,\sigma } t_{ij} c_{i\sigma}^{\dagger} 
c_{j\sigma}^{\phantom \dagger}
+ U \sum_{j} \, n_{j \uparrow}n_{j \downarrow}
-  \mu \sum_{j,\sigma} \, n_{j \sigma} ~, \label{eq:HHub}
\end{equation}
where $c_{j\sigma}$ is the annihilation operator for an electron at
site $j$ with spin $\sigma$.  
$t_{ij}$ is the nearest neighbor hopping integral, 
$U$ is the on-site repulsion between 
electrons of opposite spin, $\mu$ the chemical potential, and 
$n_{j \sigma}=c_{j \sigma}^{\dagger}c_{j \sigma}^{\phantom \dagger}$ 
is the occupation number operator.
Disorder is introduced by taking the hopping parameters $t_{ij}$ from
a probability distribution $P(t_{ij}) = 1/\Delta$ for
$t_{ij} \in [1-\Delta/2, 1+\Delta/2]$, and zero otherwise.
$\Delta$ is a measure for the strength of the disorder \cite{dishubqmc}.

We use the Determinant Quantum Monte Carlo (QMC) method, which has been 
applied extensively to the Hubbard model without disorder \cite{WhHbk}.
While disorder and interaction can be varied in a controlled way and
strong interaction is treatable, QMC is limited in the size of the 
lattice, and the {\em sign problem} restricts the temperatures 
which can be studied. The sign problem is minimized
by choosing off-diagonal rather than diagonal disorder, 
as at least at half-filling ($\langle n \rangle=1$) there is no sign 
problem in the former case, and consequently simulations can be pushed 
to significantly lower temperatures. 
For results away from half filling we choose $\langle n \rangle = 0.5$ 
where the sign problem is less severe compared to other 
densities\cite{WhHbk}. Also, interestingly, the sign problem is reduced 
in the presence of disorder \cite{dishubqmc}.

The quantity of immediate interest when studying a possible metal--insulator
transition is the {\em conductivity} and especially its $T$-dependence.
By the fluctuation--dissipation theorem $\sigma_{dc}$ is related to the
zero-frequency limit of the current-current correlation function. 
A complication of the QMC simulations is that the correlation functions 
are obtained as a function of {\em imaginary} time.
To avoid a numerical analytic continuation procedure to obtain 
frequency-dependent quantities, which would require Monte Carlo data of 
higher accuracy than produced in the present study, we employ an 
approximation that was used and tested before in studies of the
superconductor--insulator transition in the attractive Hubbard 
model \cite{Trivsig}. This approximation
is valid when the temperature is smaller  
than an appropriate energy scale in the problem.
Additional checks and applicability to the present problem 
are discussed below.
The approximation allows $\sigma_{\rm dc}$ to be computed directly from
the wavevector ${\bf q}$- and imaginary time $\tau$-dependent 
current-current correlation function $\Lambda_{xx} ({\bf q},\tau)$:
\begin{equation}
 \sigma_{\rm dc} = 
   \frac{\beta^2}{\pi} \Lambda_{xx} ({\bf q}=0,\tau=\beta/2) ~.
 \label{eq:condform}
\end{equation}
Here $\beta = 1/T$,
$\Lambda_{xx} ({\bf q},\tau) = 
   \langle j_x ({\bf q},\tau) \, j_x (-{\bf q}, 0) \rangle$,
and $j_x ({\bf q},\tau)$ the ${\bf q},\tau$-dependent current 
in the $x$-direction, is the Fourier transform of
$j_x ({\bf \ell}) =  i \sum_\sigma \, t_{{\bf \ell} + \hat{x},{\bf \ell}}
(c^{\dagger}_{{\bf \ell} + \hat{x},\sigma}
c^{\phantom \dagger}_{{\bf \ell}\sigma}
- c^{\dagger}_{{\bf \ell}\sigma}
c^{\phantom \dagger}_{{\bf \ell}+\hat{x},\sigma})~$.
(see also Ref. \cite{SWZ}).

\vskip-1.5cm
\begin{figure}
  \epsfig{figure=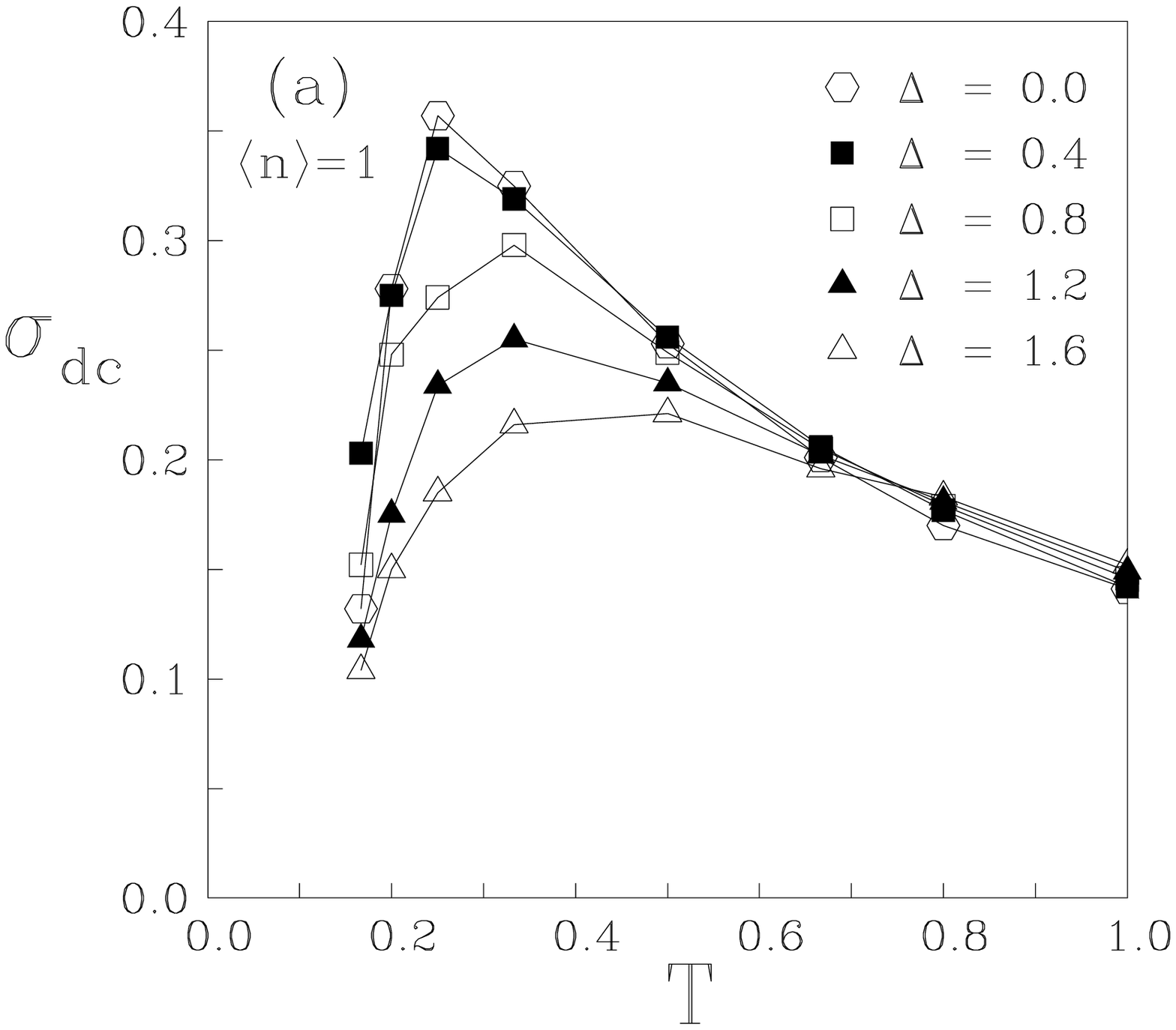,width=\linewidth}
\vskip-2.0cm
  \epsfig{figure=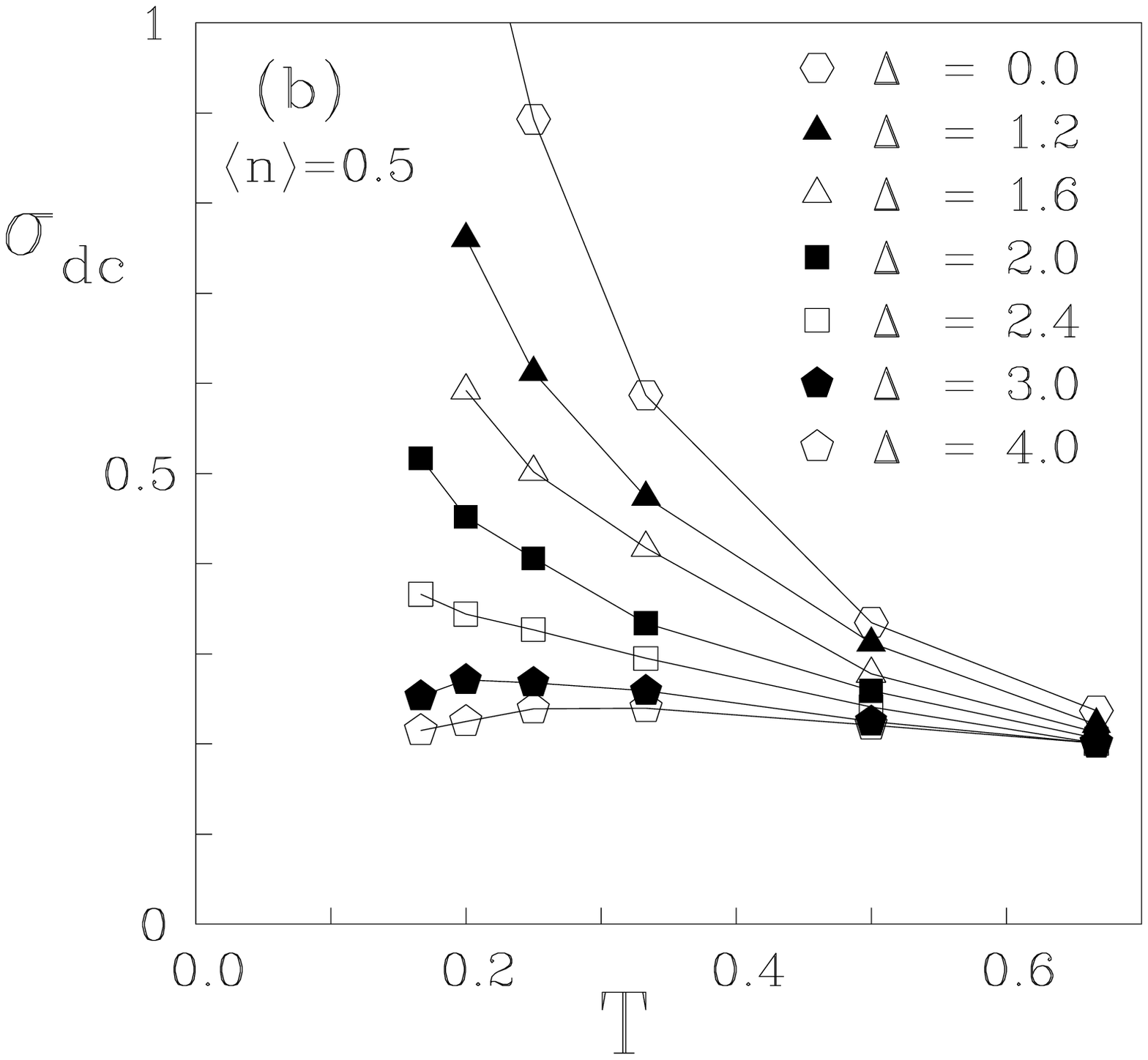,width=\linewidth}
\vskip-0.5cm
 \caption{\label{fig:Fig1} Conductivity $\sigma_{\rm dc}$ as a 
function of temperature $T$ for various values of disorder strength 
$\Delta$ at $U=4$  for (a) half-filling ($\langle n \rangle =1$) 
and (b) $\langle n \rangle=0.5$.
Calculations are performed on an $8 \times 8$ square lattice; 
data points are averages over 4 realizations 
for a given disorder strength.}
\end{figure}

As a test for our conductivity formula (\ref{eq:condform}) we first present
results in Fig.~1(a) for $\sigma_{\rm dc}(T)$ at half-filling  
for $U=4$ and various disorder strengths $\Delta$.
The behavior of the conductivity shows that as the temperature is 
lowered below a 
characteristic gap energy, the high $T$ ``metallic'' behavior crosses over 
to the expected low $T$ Mott insulating behavior for all $\Delta$, 
thereby providing a reassuring check of formula (2) and our numerics. 

In Fig.~1(b), we show $\sigma_{\rm dc}(T)$ for a range of disorder strengths
at density $\langle n \rangle = 0.5$ and $U=4$.
The figure displays a striking indication of a change from metallic behavior 
at low disorder to insulating behavior above a critical disorder
strength, $\Delta_{\rm c} \simeq 2.7$.
If this persists to $T=0$ and in the thermodynamic limit, it would 
describe a ground state metal--insulator transition driven by \hfil\break
disorder.

\vskip-1.5cm
\begin{figure}
  \epsfig{figure=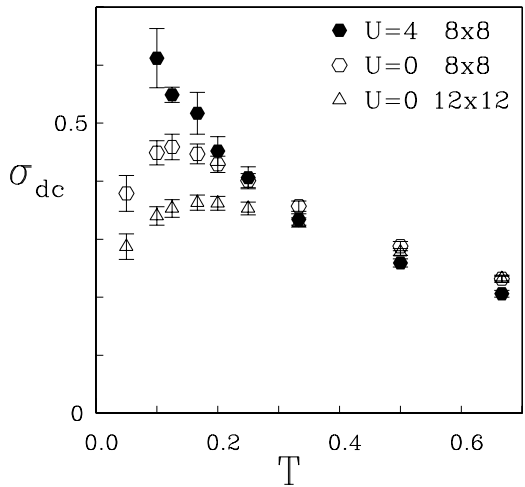,width=\linewidth}
\vskip-0.5cm
 \caption{\label{fig:Fig2} Conductivity $\sigma_{\rm dc}$ as 
a function of temperature $T$ comparing $U=4$ and $U=0$
at $\langle n \rangle=0.5$ and disorder strength $\Delta=2.0$.
Data points are averages over many realizations for this disorder 
strength (see text). Error bars are determined by the disorder 
averaging and not the Monte Carlo simulation.}
\end{figure}

In order to obtain a more precise understanding of the role of interactions 
on the conductivity, we compare in Fig.~2 the results of Fig.~1(b) with the 
disordered {\em non-interacting} $\sigma_0$ \cite{errorbars}.
The comparison is made at strong 
enough disorder $\Delta=2.0$ such that the localization length is less 
than the lattice size and the non-interacting system is 
therefore insulating with $d\sigma_0/dT >0$ at low $T$.
Interactions are found to have a profound effect on the conductivity: 
in the high-temperature ``metallic'' region, interactions 
slightly reduce $\sigma$ 
compared to the non-interacting $\sigma_0$ behavior. On the other hand 
in the low-temperature ``insulating'' region of $\sigma_0$ the data 
shows that upon turning on the Hubbard interaction 
the behavior is completely changed with $d\sigma/dT<0$, characteristic of 
metallic behavior. This is the regime of interest for the MIT.

In order to ascertain that the phase produced by repulsive 
interactions at low $T$ is not an insulating phase with a 
localization length larger than the system size
but a true metallic phase we have studied the conductivity response
for varying lattice sizes. We find a markedly different size dependence for 
the $U=0$ insulator and the $U=4$ metal, resulting in a confirmation 
of the picture given above.
For $U=0$, the conductivity on a larger ($12\times 12$) 
system is {\em lower} than that on a smaller ($8\times 8$) system 
(see Fig.~2), consistent with insulating behavior in the thermodynamic 
limit, whereas for $U=4$ the 
conductivity on the larger ($8\times 8$)
system is {\em higher} than that on the smaller ($4\times 4$) system 
(data not shown), indicative of metallic behavior. 
Thus the enhancement of the conductivity by repulsive interactions
becomes more pronounced with increased lattice size \cite{dirtyboson}.

Concerning finite-size effects for the non-interacting system we note that
at lower values of $\Delta$, where the localization length exceeds the 
lattice size, $\sigma_0$ shows ``metallic'' behavior
which is diminished upon turning on the interactions \cite{data}. 
Based on our analysis above, we would predict that at low enough $T$ and 
large enough lattice size, the conductivity curves for the non-interacting 
$\sigma_0$ and interacting $\sigma$ cross with $\sigma>\sigma_0$ at 
sufficiently low $T$.

\vskip-1.0cm
\begin{figure}
\vskip-0.5cm
  \epsfig{figure=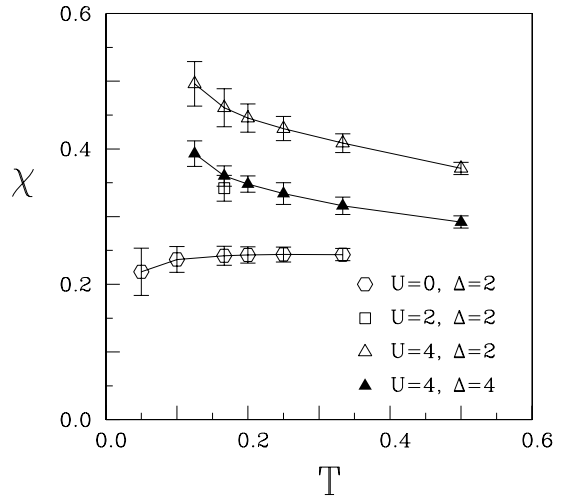,width=\linewidth}
 \caption{\label{fig:Fig3} Spin susceptibility $\chi$ as 
a function of temperature $T$ at $\langle n \rangle=0.5$ 
comparing interaction strengths $U=0, 2, 4$
and disorder strengths $\Delta=2, 4$.
Calculations are performed on $8 \times 8$ square lattices; 
error bars are from disorder averages over up to 8 realizations.}
\end{figure} 

To obtain information on the spin dynamics of the electrons 
and because it is a quantity often discussed in connection with the
localization transition, we also
compute the spin susceptibility $\chi$ as a function of T (through
$\chi(T) = \beta S_0(T)$ where $S_0$ is the magnetic structure factor
at wavevector ${\bf q} = 0$). Fig.~3 shows two things:
1) $\chi(T)$ is enhanced by interactions with respect to the non-interacting 
case (at fixed disorder strength), and 2) starts to diverge when $T$ is 
lowered, both on the metallic ($\Delta=2$) and insulating ($\Delta=4$)
sides of the alleged transition.
This is in agreement with experimental and theoretical work on
phosphorus-doped silicon, where a (3D) MIT is known to occur
and the behavior is explained by the existence of local moments\cite{SiP},
and also with diagrammatic work on 2D disordered, interacting 
systems\cite{CDC}.

In order to definitively establish the existence 
of a possible quantum phase transition 
in the disordered Hubbard model requires: 
(i) Extending the present data at $T=0.1=W/80$,
where $W$ is the non-interacting bandwidth, to lower $T$,  
which is however difficult because of the sign problem. 
(ii) A more detailed analysis of the scaling behavior in both
linear dimension and some scaled temperature.
(iii) A more accurate analytic continuation procedure 
to extract the conductivity. 
The condition for the validity of the approximate formula 
(\ref{eq:condform}) for $\sigma_{dc}(T)$, requires that
$T$ be less 
than an appropriate energy scale which is fulfilled within the two 
phases, but breaks down close to a quantum phase transition where
the energy scale vanishes. 

 
In summary, we have studied the temperature-dependent conductivity $\sigma(T)$
and spin susceptibility $\chi(T)$ of a model
for two-dimensional electrons containing both disorder and interactions.
We find that the Hubbard repulsion can enhance the conductivity and lead
to a clear change in sign of $d\sigma/dT$.  
More significantly, 
from a finite size scaling analysis
we demonstrate that repulsive
interactions can drive the system from one phase to a different phase. We find
that $\sigma(T)$ has the opposite behavior as a
function of system size in the two phases indicating that the
transition is from a localized insulating to an extended metallic phase.
The $\chi(T)$ 
data further suggests the formation of an unusual metal, a non-Fermi liquid
with local moments.
While the simplicity of the model we study prevents
any quantitative connection to recent experiments
on Si-MOSFETs, there is nevertheless an interesting
qualitative similarity between Fig.~1(b) and the experiments.
Varying the disorder strength $\Delta$ at fixed 
carrier density $\langle n \rangle$, as in our calculations, 
can be thought of as equivalent to
varying carrier density at fixed disorder strength, as in experiments,
since in a metal--insulator transition
one expects no qualitative difference between tuning the mobility edge 
through the Fermi energy (by varying $\Delta$) and {\it vice-versa}
(by varying $\langle n \rangle$).
Our work then suggests that electron-electron interaction induced
conductivity plays a key role in the 2D metal--insulator
transition.

We would like to thank C. Huscroft for useful comments on the 
manuscript, H.V. Kruis for help with the calculations,
and D. Belitz, R.N. Bhatt, C. Di Castro, T.R. Kirkpatrick, T.M. Klapwijk,
S.~V. Kravchenko, M.P. Sarachik, and G.T. Zimanyi for stimulating discussions.
Work at UCD was supported by the SDSC, by the CLC program of UCOP,
and by the LLNL Materials Institute.


\begin{references}
\bibitem{errorbars} We also consider lower temperatures and
average over larger numbers of disorder realizations: 
for the higher temperatures generally 4 realizations 
are sufficient, for the lowest temperatures the fluctuations are 
larger and up to 20 (72) realizations are used for $U=4$ ($U=0$).
\end{references}
\end{document}